\documentstyle[aps,amstex,amssymb,twocolumn,epsf]{revtex}

\begin{document}
\newcommand{\tc}{$T_{c}$ }
\newcommand{\tcy}{$T_{c}$}
\newcommand{\hcl}{$H_{c1}$ }
\newcommand{\hcly}{$H_{c1}$}
\newcommand{\ef}{$E_{F}$ }
\newcommand{\efy}{$E_{F}$}
\newcommand{\estar}{$E^{*}$ }
\newcommand{\estary}{$E^{*}$}
\newcommand{\htc}{high-temperature superconductors }
\newcommand{\htcy}{high-temperature superconductors}
\newcommand{\et}{{\it et al. }}
\newcommand{\ety}{{\it et al.}}
\newcommand{\be}{\begin{equation}}
\newcommand{\ee}{\end{equation}}
\newcommand{\hh}{$H$ }
\newcommand{\hhy}{$H$}
\newcommand{\hc}{$H_{c}$ }
\newcommand{\hcy}{$H_{c}$}
\newcommand{\ho}{$H_{0}$ }
\newcommand{\jc}{$j_{c}$ }
\newcommand{\jcy}{$j_{c}$}
\newcommand{\sg}{superconducting }
\newcommand{\sgy}{superconducting}
\newcommand{\ssc}{superconductor }
\newcommand{\sscy}{superconductor}
\newcommand{\hcu}{$H_{c2}$ }
\newcommand{\hcuy}{$H_{c2}$}
\newcommand{\rfff}{$\rho_{f}$ }
\newcommand{\rfffy}{$\rho_{f}$}
\newcommand{\hcut}{$H_{c2}(T)$ }
\newcommand{\hcuty}{$H_{c2}(T)$}
\newcommand{\jd}{$j_{o}$ }
\newcommand{\jdy}{$j_{o}$}
\newcommand{\ybco}{Y$_{1}$Ba$_{2}$Cu$_{3}$O$_{7-\delta}$ }
\newcommand{\ybcoy}{Y$_{1}$Ba$_{2}$Cu$_{3}$O$_{7-\delta}$}
\newcommand{\lsco}{La$_{2-x}$Sr$_{x}$CuO$_{4}$ }
\newcommand{\lscoy}{La$_{2-x}$Sr$_{x}$CuO$_{4}$}
\newcommand{\rrhon}{$\rho_{n}$ }
\newcommand{\rrhony}{$\rho_{n}$}
\newcommand{\rrho}{{$\rho$} }
\newcommand{\rrhoy}{{$\rho$}}
\newcommand{\qp}{quasiparticle }
\newcommand{\qpy}{quasiparticle}
\newcommand{\qps}{quasiparticles }
\newcommand{\qpsy}{quasiparticles}
\newcommand{\bib}{\bibitem}
\newcommand{\ib}{{\em ibid. }}
\newcommand{\taue}{$\tau_{\epsilon}$ }
\newcommand{\tauey}{$\tau_{\epsilon}$}
\newcommand{\vstary}{$v^{*}$}
\newcommand{\vstar}{$v^{*}$ }
\newcommand{\rhostar}{$\rho^{*}$ }
\newcommand{\rhostary}{$\rho^{*}$}
\newcommand{\vinf}{$v_{\infty}$ }
\newcommand{\vinfy}{$v_{\infty}$}
\newcommand{\fd}{$F_{d}$ }
\newcommand{\fdy}{$F_{d}$}
\newcommand{\fe}{$F_{e}$ }
\newcommand{\fey}{$F_{e}$}
\newcommand{\fl}{$F_{L}$ }
\newcommand{\fly}{$F_{L}$}
\newcommand{\jstar}{$j^{*}$ }
\newcommand{\jstary}{$j^{*}$}
\newcommand{\je}{$j(E)$ }
\newcommand{\jey}{$j(E)$}
\newcommand{\vphi}{$v_{\phi}$ }
\newcommand{\vphiy}{$v_{\phi}$}
\newcommand{\blo}{$B_{1}$ }
\newcommand{\bloy}{$B_{1}$}
\newcommand{\bhi}{$B_{\infty}$ }
\newcommand{\bhiy}{$B_{\infty}$}
\newcommand{\vlo}{$v_{1}$ }
\newcommand{\vloy}{$v_{1}$}
\newcommand{\bo}{$B_{o}$ }
\newcommand{\boy}{$B_{o}$}
\newcommand{\eo}{$E_{o}$ }
\newcommand{\eoy}{$E_{o}$}

\title{Steps in the negative-differential-conductivity regime of a
superconductor}

\author{Milind N.\ Kunchur $\dagger$$^{*}$, B. I.\ Ivlev $\dagger\ddagger$,
and J. M. Knight $\dagger$}

\address
{$\dagger$Department of Physics and Astronomy\\
University of South Carolina, Columbia, SC 29208\\
$^{*}$http://www.cosm.sc.edu/kunchur\\
and\\
$\ddagger$Instituto de F\'{\i}sica, Universidad Aut\'onoma de San Luis Potos\'{\i}\\
San Luis Potos\'{\i}, S. L. P. 78000 Mexico}
\maketitle

\begin{abstract}
\begin{sloppypar}
Current-voltage characteristics were measured in the mixed state of
Y$_{1}$Ba$_{2}$Cu$_{3}$O$_{7-\delta}$
 superconducting films in the regime where flux flow becomes
unstable and the differential conductivity $dj/dE$ becomes negative.
Under conditions where its negative slope is steep,
the $j(E)$ curve develops a pronounced staircase like
pattern. We attribute the steps in $j(E)$ to the formation of a
dynamical phase consisting of the successive nucleation of
quantized distortions in the local vortex velocity and
flux distribution within the moving flux matter.
\end{sloppypar}
\end{abstract}


\narrowtext
In a type II superconductor,
a magnetic field $H$ above the lower critical value \hcl
introduces
flux vortices  containing an elementary quantum of
flux $\Phi_{o}=h/2e$, and interactions between the vortices
tend to align them into a uniform lattice \cite{abrikosov}. Extrinsic
forces due to impurity pinning, thermal fluctuations, or dynamic melting
can result in a disordered solid or liquid state instead of a
crystalline lattice \cite{blatter}; however, long-range repulsions
between the vortices will still enforce a relatively uniform
density. The system under study consists of a superconducting film in
a perpendicular applied flux density \bo along $\hat{z}$, with a transport
electric current density $j$ and electric field $E$ along $\hat{y}$
in the plane of the film. The transverse component of $E$ is negligible
for this discussion and the vortices move with velocity \vphi
predominantly along $\hat{x}$.

A transport current exerts a Lorentz driving force
$\bf F_{L} = j \times \Phi_{o}$
on the vortices and the motion is opposed by a viscous drag
$\bf F_{d}$$= -\eta$$\bf v_{\phi}$, where $\eta$ is the coefficient of
viscosity.
If we assume that pinning forces $\bf F_{p}$ are negligible
(because $F_{L} \gg F_{p}$) then the steady state
motion reflects a balance between
driving ({\bf \fly }), drag ({\bf \fdy }), and elastic forces ({\bf \fey })
on each vortex.
For a perfectly uniform distribution, the net elastic force on each
vortex vanishes resulting in free flux flow \cite{obs}. Then
$j \Phi_{o} =  \eta v_{\phi}$ producing an Ohmic response
approximated by \cite{blatter,bs}:
${\rho_{f}}/{\rho_{n}} \approx {B}/{H_{c2}(T)}$.

A different scenario prevails at
ultra high dissipation levels and electric fields sufficient to
alter the electronic distribution function and/or the electronic temperature.
Here $j(E)$ becomes non-linear and can develop an unstable region with
negative differential conductivity (NDC) (region ``C''
in Fig.~1). Such unstable behavior and NDC has been predicted by
Larkin and Ovchinnikov (LO) for the regime near \tc \cite{LO}, and has
been experimentally well established \cite{klein,loexpt}. In the LO
mechanism, a non-equilibrium electronic distribution function leads to
shrinkage of the vortex core by removal of
quasiparticles from its vicinity \cite{LO,klein}.
In the opposite regime of
$T \ll T_{c}$ we have observed a qualitatively different type of instability
 \cite{loTinst,aps01} that seems to result from a
temperature differential between the electronic system
and lattice while maintaining an equilibrium-like distribution function
(because electron-electron scattering is more rapid than
electron-phonon scattering at low $T$).
In our low-temperature instability the vortex expands rather than
shrinks, and viscous drag is reduced because of a softening of
gradients of the vortex profile rather than a removal of
quasiparticles. The standard LO instability
near \tc occurs at values \jstar and \vstary =\estary/$B$ (see Fig.~1)
that are $B$ independent, whereas in
our low-temperature instability \jstar and \vstar have a $\sim 1/\sqrt{B}$
dependence. This plays an important role in the appearance of a staircase
at low temperatures but not at the higher temperatures of the previously
well studied LO phenomenon. Theoretical and experimental details of the
low-temperature instability are discussed elsewhere \cite{loTinst,normalstate}.

Fig.~1 suggests a qualitative picture of how the drag component
($\eta v$) of the \je response
might vary with $E$ over its entire range.
After the first ``hump'' (regions A, B, and C)
because of a reduction in $\eta$ by the
mechanisms mentioned above, \je will rise again
when \vphi reaches some limiting value
such as $v_{\infty} \sim \xi/\tau_{\Delta}$,
where $\xi = 1.5$ nm is the
coherence length and $\tau_{\Delta} = \hbar/\Delta \approx 4.7 \times
10^{-14} s$ is the
order-parameter relaxation time. If the vortex density is non-uniform, an
additional elastic term appears in $j$, namely $j =(\eta v - F_{e})/\Phi_{o}$.
We shall refer to the \je curve of Fig.~1 as the {\it primitive} curve,
which applies to a hypothetical ensemble of uniformly packed vortices
moving with identical velocities; an actual \je will not follow this behavior
since, upon encountering a negative slope, the flux matter will
become unstable and
undergo a phase transition, such that at any given time there exist
multiple vortex velocities (resulting in a composite
response) and a non-uniform vortex density (resulting in elastic corrections).
In effect some vortices will be travelling on the second positive slope ``E''
of Fig.~1 with \vphi $\sim$ \vinf (in regions with reduced vortex
density) and the rest will be travelling
at a reduced velocity \vlo $<$ \vstar  resulting in
restabilization. Below we calculate the expected composite response for
one particular flux structure and
show that it leads to a staircase in a natural way in rough
agreement with the data.

The samples are {\em c}-axis oriented epitaxial films of
\ybco on (100) LaAlO$_{3}$ substrates with \tcy 's around 90K and of
thickness $t \approx 90$ nm.
Electron-beam and optical-projection lithographies, together with
wet etching in~$\sim$~1~\%~phosphoric acid,
were used to pattern bridges of widths $w \approx$ 4 $\mu$m and lengths
$l \approx$ 90 $\mu$m. At the end of the lengthy fabrication, each microbridge
is inspected by a variety of high-resolution optical probes for
uniformity of width to ensure high reproducibility.
Altogether ten samples were studied at 12 temperatures
(1.6, 2.2, 6, 7, 8, 10, 20, 27, 35, 42, 50, 80 K) and at 11 flux densities
(0.1, 0.2, 0.5, 1, 1.5, 2, 10, 11, 13, 13.5, 14 T).
We always observe an instability with steps in the NDC region for
all temperatures below \tcy /2 and for $B$ values in the 1--14 T range.
The electrical transport
measurements were made with a pulsed constant voltage source,
preamplifier circuitry, and a digital storage oscilloscope.
The pulse rise times are about 100 ns with a duty cycle of about 1 ppm,
resulting in effective thermal resistances of order 1 nK.cm$^{3}$/W.
The ability to hold the voltage constant across the
sample allows investigation of the $j(E)$ curve in the NDC region;
however, once $dj/dE$ becomes negative there will be a jump in the
voltage until the chordal resistance
of the sample has risen above the source impedance of the
voltage source (including the resistance of the leads and contacts)
allowing a stable steady state. Such forbidden gaps in $E$
are visible in the last data plot at the onset of NDC.
Each $j(E)$ curve typically consists of 1000 separate points and each
point requires averaging over several hundred pulses to obtain an
adequate signal-to-noise ratio (SNR), so that a single curve takes
several days to measure. Note that the $j$ values in the experiment
are an order of magnitude lower than the depairing current
density \cite{pair}. Further details about the experimental techniques are
discussed elsewhere \cite{normalstate,mplb}.

Fig.~2 shows some examples of measured staircase patterns in the NDC regime.
We found such patterns to be ubiquitous under most
conditions of \bo and $T$ (not close to \tcy ).
Fig.~2(a) shows the behavior at 11T and 27K.
The main step features are seen to be reproducible between curves
measured a week apart in opposite directions of
changing $E$. Panel (b) shows 50-K  1-T \je curves
for two different samples (sample X was measured 23 days after sample Z).
Within the scatter and uncertainity in the absolute $j$ values (due to
uncertainity in the sample widths used to calculate $j$ from $I$),
there apears to be some consistency in the main step features.
The steepest NDC and most pronounced staircase patterns were
observed at temperatures around one half of \tc and flux densities around
1 Tesla (where
the intervortex separation $a \simeq \sqrt{\Phi_{o}/B}$ becomes comparable
to the penetration depth).
Fig.~2(c) shows 50K curves for sample X at three $B$ values over an
extended range of $E$. The solid line fit indicates linearity in $1/E$
over the steep NDC portion. Other observed features are a convergence of
curves in their NDC regions for different \boy 's  (at fixed $T$) and a
growth in horizontal step size with increasing $E$.

Let us now consider the nature of the transition in the flux
matter when the imposed electric field $E_{o}$ exceeds \estar and
enters the NDC region ``C'' of the primitive curve (Fig.~1). Since all
the vortices could not be moving
uniformly with \vphi = $E_{o}/B_{o}$ (for they would all be unstable)
the flux matter must reorganize itself
by the creation of moving defects in the regular vortex structure.
The exact nature of the defects will depend on the vortex state.
In a vortex solid there can be pairs of edge dislocations of different
signs, or defect-interstitial pairs that conserve the total flux. If the
moving vortex system is a liquid, the defect can be
 a finite magnetic spot or domain with a different flux density, since
redistribution of a magnetic flux in a vortex liquid does not
 involve significant energy barriers \cite{chakravarty}.
In this Letter we present a simple calculation based on the magnetic domain
approach, in order to provide a simple physical explanation of the
effect.  A detailed and more rigorous approach, which considers the
distinction among the different possibilities listed above, will be
presented elsewhere. In the reorganized flux structure some vortices at any
given time will move on the second positive slope ``E'' with \vphi $\sim$
\vinf while others move on the first slope ``A'' with  $v_{1}
<$ \vstary , such that the composite macroscopic electric field across a
sample length averages to \eoy . The reorganized flux structure is influenced
by several conditions:
(1) The ends of the sample are at a fixed potential difference so
the average electric field along any longitudinal path is close to
the applied
\eoy ; (2) the current density integrated along any longitudinal path
is approximately constant due to phase coherence (allowing for small
variations in the vector potential);
(3) the average flux density in the sample must equal the applied value \boy ,
otherwise the demagnetizing fields would diverge --- this means that if
defects are formed that have a suppressed local $B$, the flux density \blo
in the outside bulk will increase to conserve the total flux so as to
maintain $<B> =B_{0}$;
(4) extended low $B$ defects will tend to be stationary to reduce
circulating electric fields around them, which would violate the first
condition;
(5) elastic forces between adjacent vortices --- along with asymmetry
in demagnetization (since $l \gg w$) --- will tend to discourage
relative side-by-side motion and promote modulations in \vphi and $B$ that
traverse the entire length of the sample; and
(6) bulk elastic forces within the vortex matter will discourage continuous
growth of defects but instead favour
discontinuous nucleation of uniformly spaced defects of the smallest
possible size ($\sim$ 1 lattice constant).

Considering the above guidelines, we estimate below the composite \je
response for one simplified static scenario of the
defect structure. We would like to emphasize that this is not presented
as a comprehensive or unique solution to the vortex reorganization
problem, but only used to provide a simple physical explanation of the
observed staircase behavior. Referring to Fig.~1,
let the applied electric field just exceed the peak value \estary .
 The functional form of the primitive \je curve in the peak region is
given by \cite{loTinst,normalstate}
$j \sim (\sigma_{n}H_{c2}/B) E/[1 + (E/E^{*})^{2}]$ with
$E^{*}=\sqrt{2\rho_{n}Bn\Delta/(H_{c2}\tau_{\epsilon})}$ and
$j^{*}= \sqrt{n\Delta H_{c2}/(2\rho_{n}B\tau_{\epsilon})}$.
Some vortices are forced to move at higher velocity in order to
restore stability.
For the purpose of obtaining a rough estimate,
let's assume that vortices move at \vinf when they cross a
longitudinal channel with reduced local $B$ and
enlarged flux lattice spacing $A$
(The actual structure may consist of an array of smaller entities
that nucleate at the sample edges and migrate inward.)
The smallest width
of this channel, consistent with continuity of flux and constancy of
longitudinal $E$, is a single lattice spacing $A= a (v_{\infty}$/\vloy ),
where $a$ and \vlo are the lattice spacing and velocity
of the ``bulk'' vortex matter outside the channels.
As soon as the first defect is nucleated, the flux density \blo in the
bulk increases discontinuously  \footnote{These estimates neglect the relatively small flux\newline(\bloy $Al$\vloy\vinfy) through the defects, which, however, is
included in the numerically calculated curves of Fig.~3(b).}
from \bo to \boy$w/(w-A)$ and \vlo
drops from \eoy/\bo to \eoy ($w-A$)/(\boy $w$).
This immediately
stabilizes flux flow in the bulk since
\estar $\propto \sqrt{B}$ increases and these vortices
drop back on the positive slope ``A'' (in Fig.~1).
As the applied \eo is increased further nothing more happens
until \vlo reaches the new peak value \estary /\bloy .
 Then a second defect is nucleated and the process keeps repeating itself
until the defects fill about half the sample area, at which point it
becomes energetically favourable to continuously expand existing defects.
The top of each step corresponds to $j=$\jstary (\bloy) for each new \blo.
Since \jstar $\propto 1/\sqrt{B}$ and \estar $\propto \sqrt{B}$,
the staircase shows an overall downward trend with $j \sim 1/E$
dependence that does not depend on \bo (causing curves with different
applied \bo to converge). Both of
these behaviors can be seen in Fig.~2(c). Fig.~3 shows experimental and
numerically computed \je curves at $T$=50K and $B$=1, 1.5, and 2 Tesla.
The two sets of curves show similar qualitative trends and have roughly
comparable magnitudes. Note: The curves computed from the model
have no adjustable parameters --- the values for
$1/\sigma_{n} \simeq \rho_{n} = 57 \mu \Omega$-cm and
$H_{c2} = 120$ T come from the literature \cite{normalstate,nakagawa}
and \estar = 35 V/cm (at one value of $B=1.5$ T) is taken from the measurement.

To summarize, we have investigated the transport response
of a superconductor into the regime of negative differential conductivity
beyond the low-temperature instability. A qualitatively new
behavior was observed in the form of steps in the \je curve.
The observed behavior is consistent with the restabilization of
a moving vortex distribution by the
formation of a dynamical phase with distortions in the local flux density and
vortex velocity. While we do not undertake a detailed
theoretical investigation of the exact nature of the dynamical phase, we
hope our experimental results will stimulate such further work.
Recently other interesting but separate effects have been seen in the
highly driven vortex state by the Huebener group, which include
time dependent oscillations and
hysteretic steps resulting from tunneling between
neighbouring vortices\cite{huebpapers}.
The steps in our effect are time-independent and
non-hysteretic, and are analogous to the
Gunn effect in semiconductors, where electric-charge
modulations lead to steps in \je in the NDC regime.

The authors acknowlege useful discussions and other assistance from
D. K. Christen, J. M. Phillips, M. Geller, A. Koshelev, R. P. Huebener,
N. Schopohl, J. Blatter, and V. Geshkenbein.
This work was supported by the U. S. Department of Energy through grant
number DE-FG02-99ER45763.

\begin{figure}[p]
\hspace*{8mm}
\vspace{2cm}
\begin{center}
\epsfxsize=0.4\textwidth\epsfbox{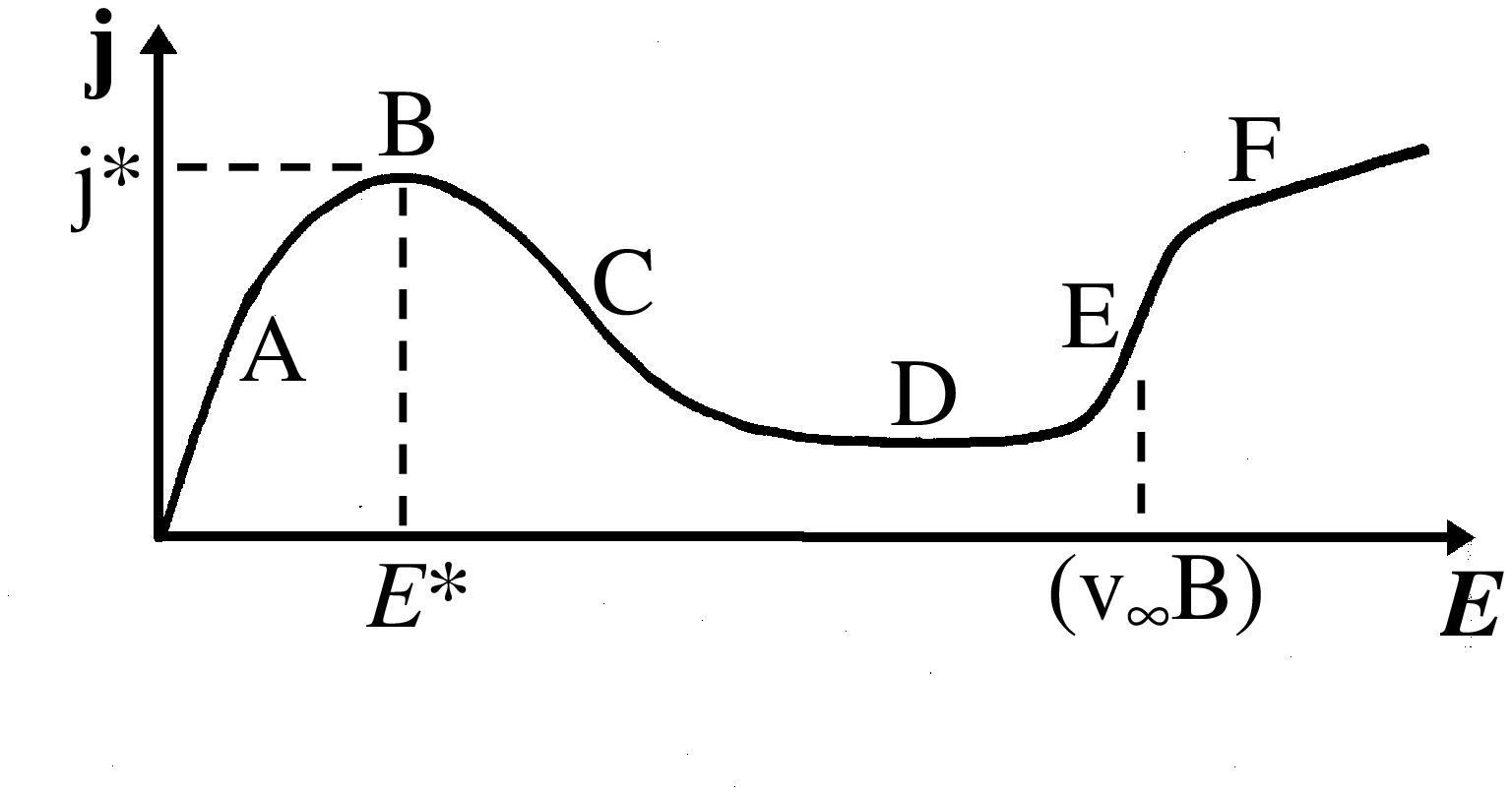}
\vspace{2cm}
\caption{Schematic diagram for a primitive \je curve for moving flux vortices distributed with uniform density and moving with the same velocity.
The net elastic force \fe on each vortex vanishes and the Lorentz
driving force \fl is balanced by the viscous drag \fd so that
$j =\eta v/\Phi_{o}$. Experimentally only the regime ``A'' is accessible.
Beyond the peak at ``B'', unstable dynamics
destroys the perfect spatial homogeneity of the vortex distribution and
leads to non-vanishing elastic forces so that $j =(\eta v -
F_{e})/\Phi_{o}$. Now vortices travel in differing elastic environments
and at unequal velocities such that each vortex is traveling on
either positive slope ``A'' or ``E'' on the primitive \je curve appropriate
to its local instantaneous enviroment. The positive
slope ``E'' arises upon reaching some limiting
velocity $v_{\infty} < \xi/\tau_{\Delta}$ and ``F'' is entered when the
sample is driven normal.}
\end{center}
\label{Fig. 1}
\end{figure}

\begin{figure}[p]
\hspace*{8mm}
\vspace{1cm}
\begin{center}
\epsfxsize=0.4\textwidth\epsfbox{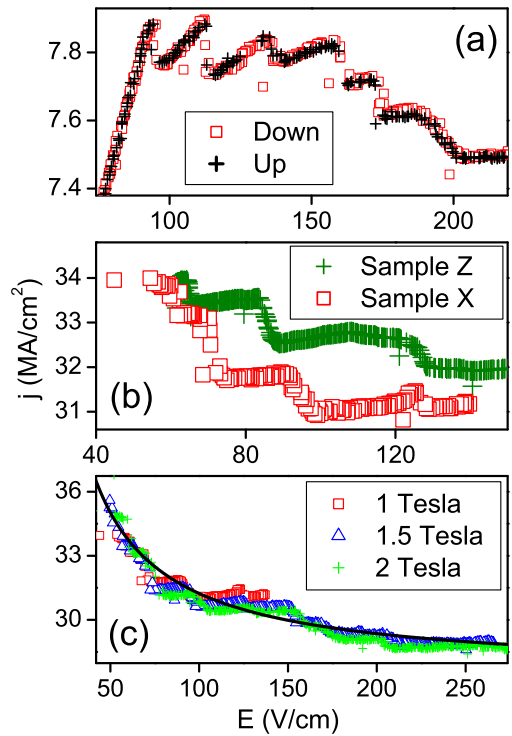}
\vspace{2cm}
\caption{(a) Experimental \je curves for sample X
at \boy =11T and $T$=27 K.
The two curve sets were measured in increasing (up) and decreasing
(down) $E$ to confirm reproducibility and absence of hysteresis.
(b) Magnified view of experimental \je curves
for two different samples at 50 K and 1 T (sample X was measured 23 days after sample Z).
The step features show some consistency within the scatter of the data and
uncertainities in the two sample widths used to calculate $j$.
(c) 50K low-$B$ \je curves for sample X over an extended
$E$ range showing overall linearity in $1/E$; the solid line represents
$j=27 + 375/E$..}
\end{center}
\label{Fig. 2}
\end{figure}

\begin{figure}[p]
\hspace*{8mm}
\vspace{2cm}
\begin{center}
\epsfxsize=0.4\textwidth\epsfbox{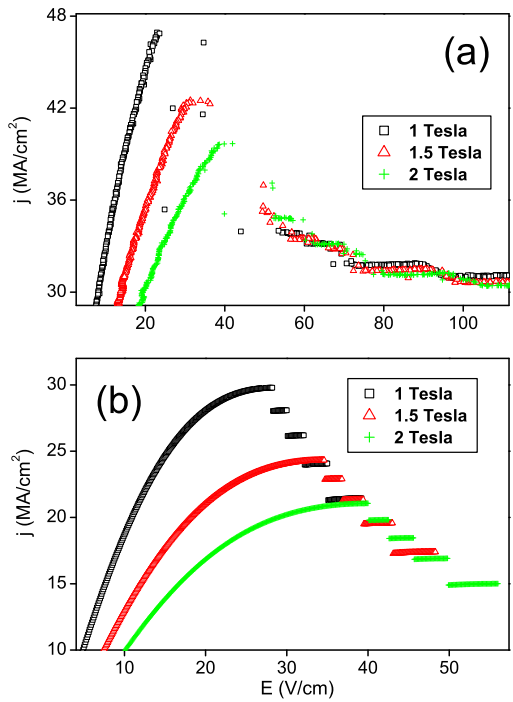}
\vspace{2cm}
\caption{ (a) Experimental \je curves for sample X at $T$=50 K and $B$= 1,
1.5, and 2 T. In their NDC regimes, the curves collapse onto an
approximately common behavior, where $j$ is roughly linear in $1/E$.
There are forbidden gaps in $E$ just beyond the peak,
when the sample's resistance is less than the voltage source impedance,
as discussed in the text.
(b) Theoretical \je curves for the
same $T$ and $B$'s, calculated from the simple model based on dynamical
flux reorganization (the model has no fitting parameters).}
\end{center}
\label{Fig. 3}
\end{figure}

\end{document}